\documentclass[11pt,twoside]{article}
\usepackage{asp2010}

\resetcounters

\begin{document}

\title{Some recent results from the CHARA Array}
\author{T.A.~ten~Brummelaar$^1$, D.~Huber$^2$, K.~von~Braun$^3$, \
	T.~Boyajian$^4$, N.D.~Richardson$^4$, G.~Schaefer$^1$, \
	I.~Tallon-Bosc$^5$, D.~Mourard$^6$, H.A.~McAlister$^4$,\
	N.H.~Turner$^1$, L.~Sturmann$^1$, J.~Sturmann$^1$, J.D.~Monnier$^8$\
	and M.~Ireland$^9$
\affil{$^1$The CHARA-Array of Georgia State University, Mt Wilson CA}
\affil{$^2$The University of Sydney, N.S.W. 2006 Australia}
\affil{$^3$NExScI / California Institute of Technology, Pasadena CA}
\affil{$^4$CHARA Georgia State University, Atlanta GA}
\affil{$^5$CNRS, UMR 5574, Centre de Recherche Astrophysique de Lyon, 9 avenue\
Charles Andr\'e, Saint-Genis Laval, F-69230, France}
\affil{$^6$Obs de la C\^ote d'Azur, Nice France}
\affil{$^7$National Optical Astronomonical Observatory - Tucson AZ}
\affil{$^8$University of Michigan, Ann Arbor, MI}
\affil{$^9$Macquarie University, NSW 2109 Australia}}

\begin{abstract}
The CHARA Array is a six 1-m telescope optical and near infrared 
interferometer located at the Mount Wilson Observatory in southern 
California and operated by the Center for High Angular Resolution Astronomy 
of Georgia State University. 
The CHARA Array has been in regular scientific 
operation since 2005 and now has over 55 publications in the refereed 
literature, including two in Science and one in Nature. The Array now 
supports seven beam combiners ranging from 0.5 microns up to 2.3 microns 
and combing from 2 to 4 beams at a time. An upgrade to a full 6 beam combiner 
is now underway and fringes with all six telescopes were achieved soon 
after the meeting.
We present some of the more recent results from the CHARA-Array.
\end{abstract}

\section{Introduction}
The CHARA Array \citep{CHARA05} has been in regular scientific operation 
since 2005. Many recently published results are to be found elsewhere in these 
proceedings 
\citep{Aufdenberg2011, Baron2011, Botajian2011, Kloppenborg2011, Monnier2011,
Parks2011, Rajagopal2011, Gail2011, Simon2011, Stencel2011, Touhami2011,
White2011, Zhao2011}, and so in this paper we will focus on work
in progress that had not been published at the time of this meeting.

\section{NOAO Time}
Applications for observing time in 2010 and 2011 were solicited and
processed by the 2010A and 2011A NOAO proposal cycles. During 2010, 
7.4 nights were reserved and 13 proposals requesting 17.1 nights were received 
for an over-subscription rate of 2.31. The 2011A cycle saw 20 CHARA proposals
submitted requesting 24.5 nights, which compared with the 5.0 nights set 
aside yields an over-subscription rate of 4.90. The growth rate of proposals 
exceeded 50\% from the first to the second year. 
We regard this experience as 
providing solid evidence of strong community interest in, as well as for
the broad applicability of interferometry. 

\section{Work in Progress}

Here we describe four projects that were underway at the time of this meeting.
Since that time, two of them have been published \citep{Trinity11, Kaspar11}.

\subsection{Combining Kepler Photometry with CHARA Interferometry}

The high-precision photometry of the Kepler space telescope is currently 
delivering ground-breaking discoveries in many fields of stellar physics. 
Interferometric follow-up, however, is limited by the small sizes of the stars 
and requires the long-baselines and high sensitivity in the visible only 
available at CHARA.  See Figure \ref{fig_Trinity}.
\articlefiguretwo{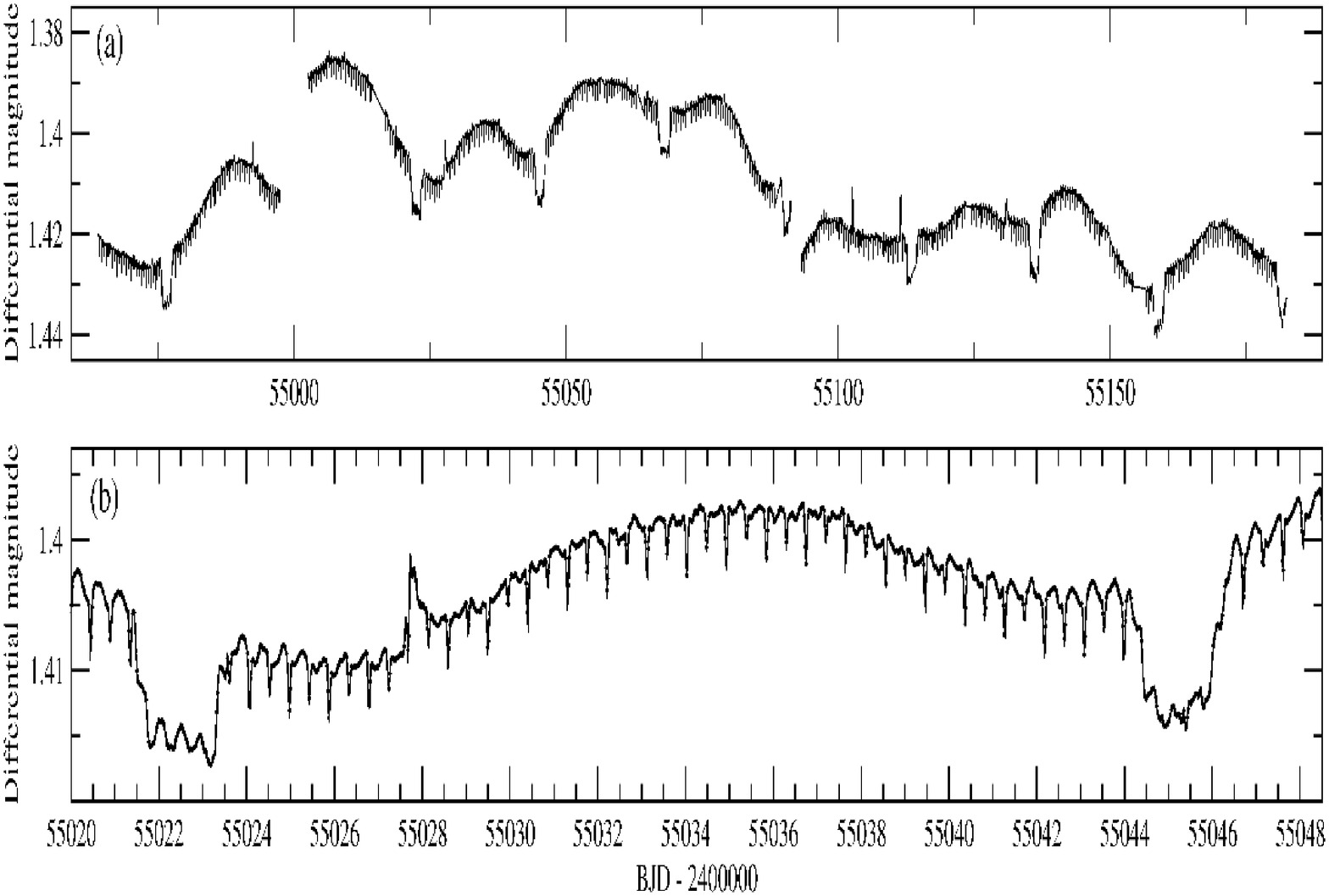}{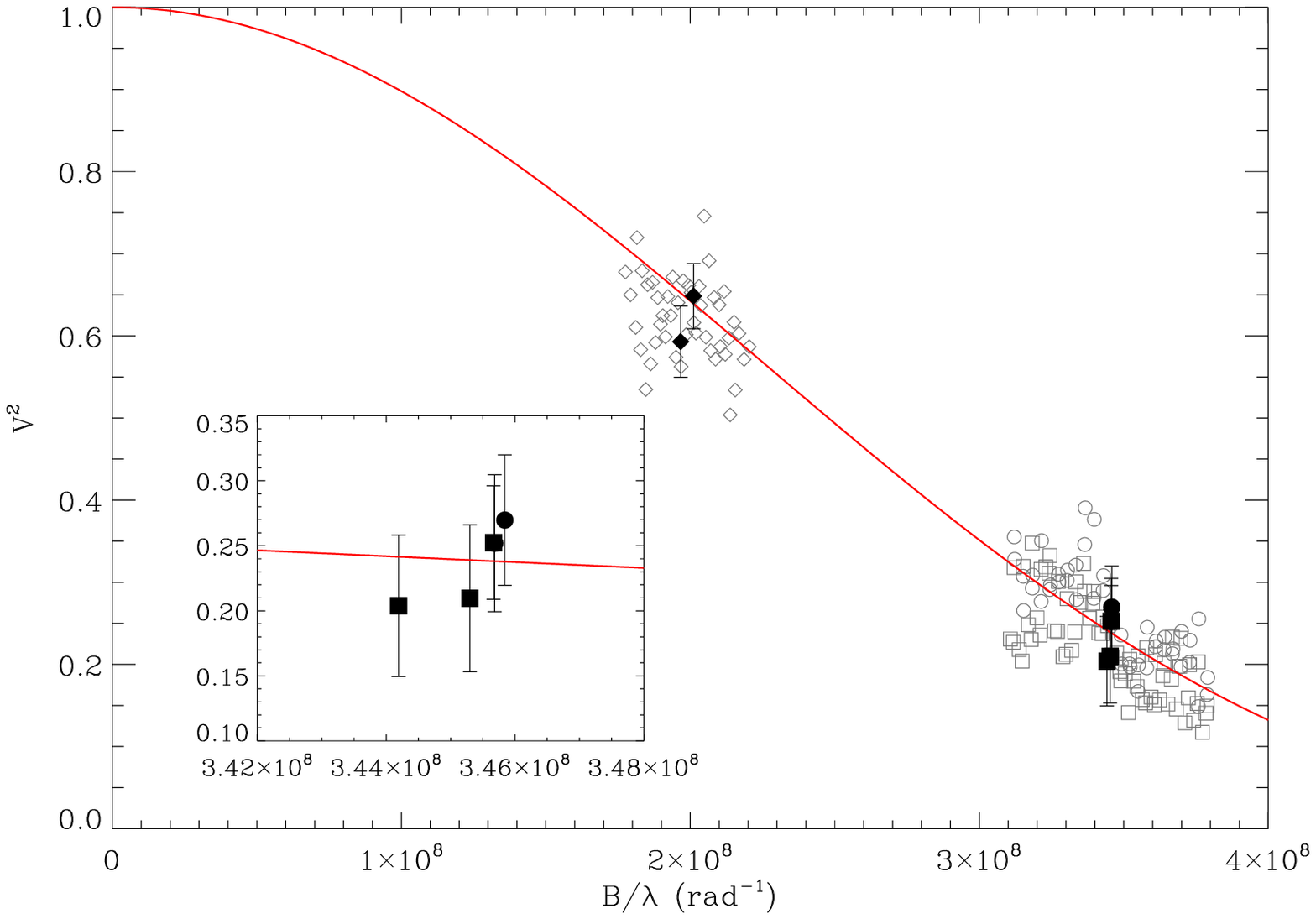}{fig_Trinity}{\
Left: The Kepler discovery of the hierarchical triple system with triple
eclipses HD 181068. As confirmed by supplementary radial-velocity data, the
primary component is eclipsed with an orbital period of 45.5 days by
a pair of faint companions which themselves eclipse each other with
an orbital period of 0.9 days.  Right: Using the PAVO beam combiner we
have measured the angular diameter of the primary to be $0.461 \pm 0.011$ mas,
which combined with the parallax yields $12.4 \pm 1.3$ Rsun. This
confirmed that the primary is a $\approx$ 3 Msun red giant orbited by pair of
red dwarfs. All three stars have very similar surface temperatures of
$\approx$ 5100K, explaining the disappearance of the 0.9 d eclipses when the
short-period binary is in front of the red giant (see lower left figure).
Further studies of this system are expected to yield important constraints on
the dynamical evolution of multiple systems.}

\subsection{The Luminous Blue Variable P Cygni}

P~Cygni is a bright (V=4.8\,mag) luminous blue variable whose well-studied
wind serves as the prototype for objects of its type.
%, but it has not yet 
%been studied at high angular reoslution in the near Infra-Red.
Simultaneous attainment of high spatial and spectral resolutions 
are enabling significant progress on constraining wind geometry 
and pinpointing the origin of the emission line flux. 
Full two-dimensional imaging is also being pursued. See Figure \ref{fig_PCyg}.
\articlefiguretwo{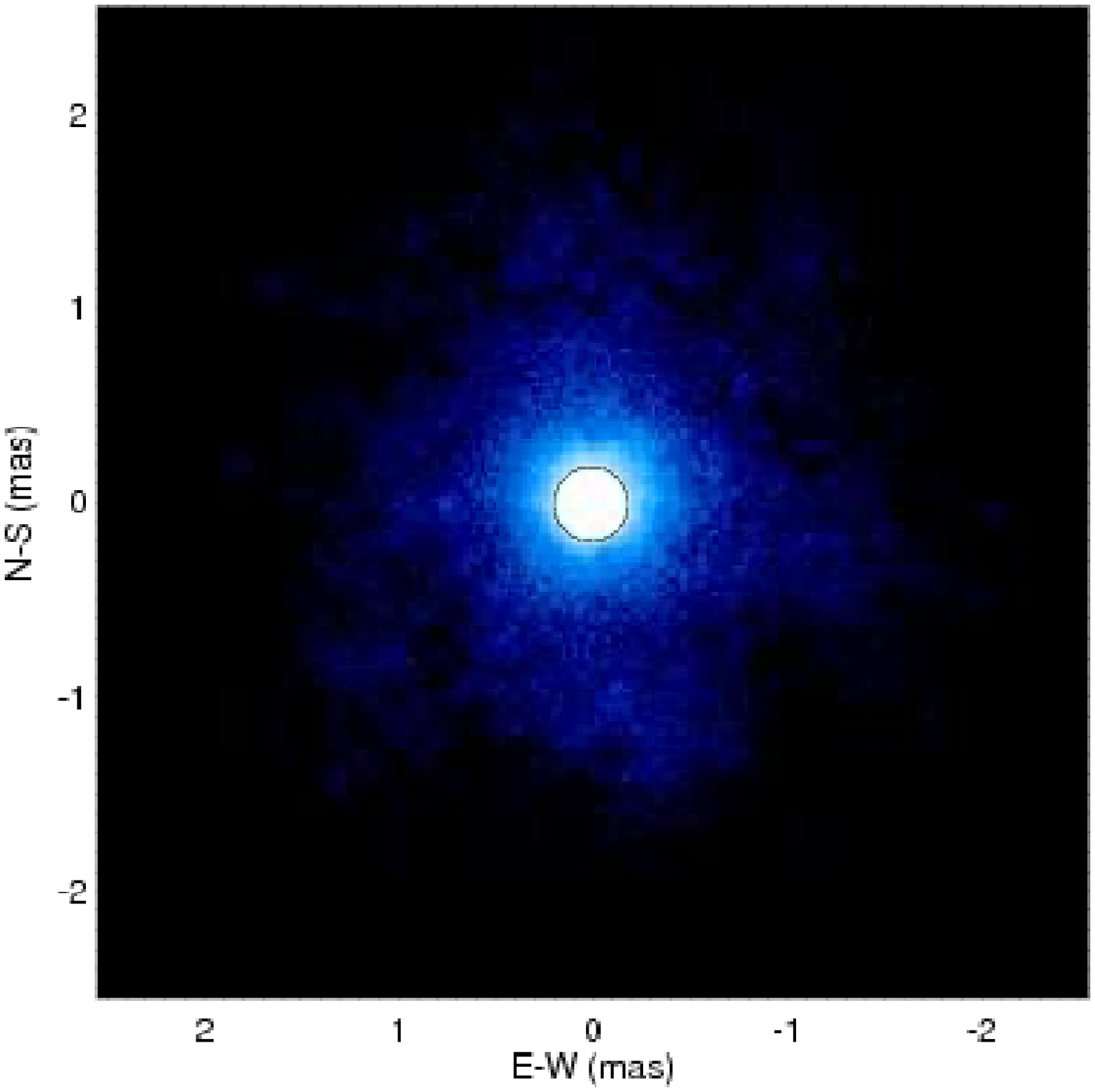}{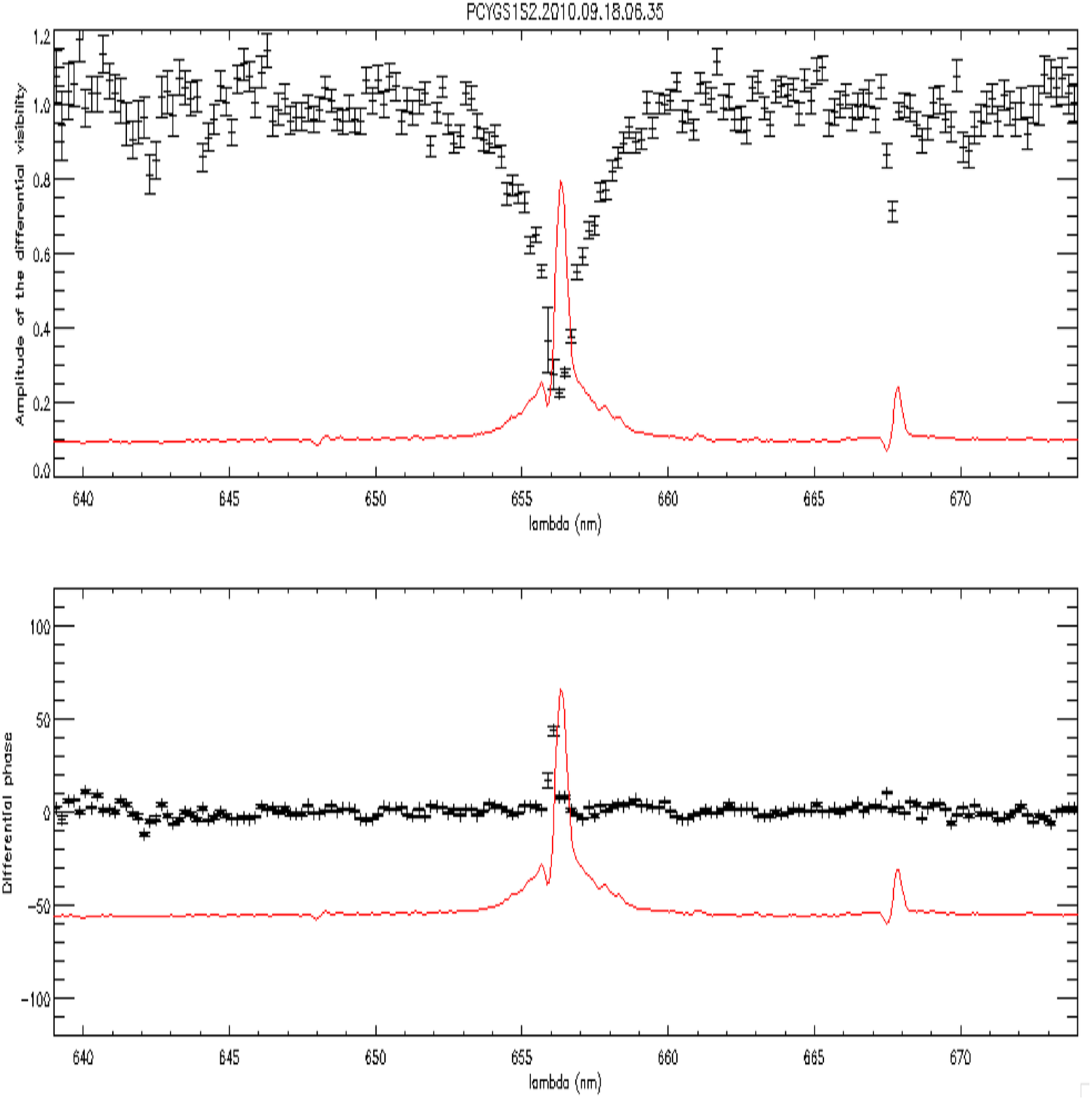}{fig_PCyg}{\
Left: A very preliminary image of P Cygni in the H band made using the\
MIRC beam combiner. The slight asymmetry is extremely similar to the \
asymmetries observed in the large scale (arc-second and arc-minute scale) \
ejecta surrounding the star. Right: The visibility (top) and \
differential phase measurements (bottom) for the smallest 30m baseline near \
the H$\alpha$ line using the VEGA beam combiner. The second smaller line
shown is HeI line. The H$\beta$ line was also monitored in a separate
spectral channel of the spectrograph and is not shown here.
The visibility \
is lower inside the lines, indicating that we are resolving the emission \
line region. There is also a clear differential phase signal in the \
line indicating strong asymmetries in the ejecta.}

\subsection{The Habitable Zone of the Exoplanet System GJ 581}

The nearby, low-mass star GJ 581 is reported to have either four or six planets 
in two rivaling scenarios in the literature. Here we directly measure GJ 581's 
surface temperature and stellar diameter, which is about three times the 
diameter of Jupiter. A direct consequence identifying the location and 
extent of GJ 581's Habitable Zone (HZ). Figure \ref{fig_GJ581} 
shows the location of 
GJ 581's HZ (gray shaded region) and its planets in the two literature 
scenarios. 
\articlefiguretwo{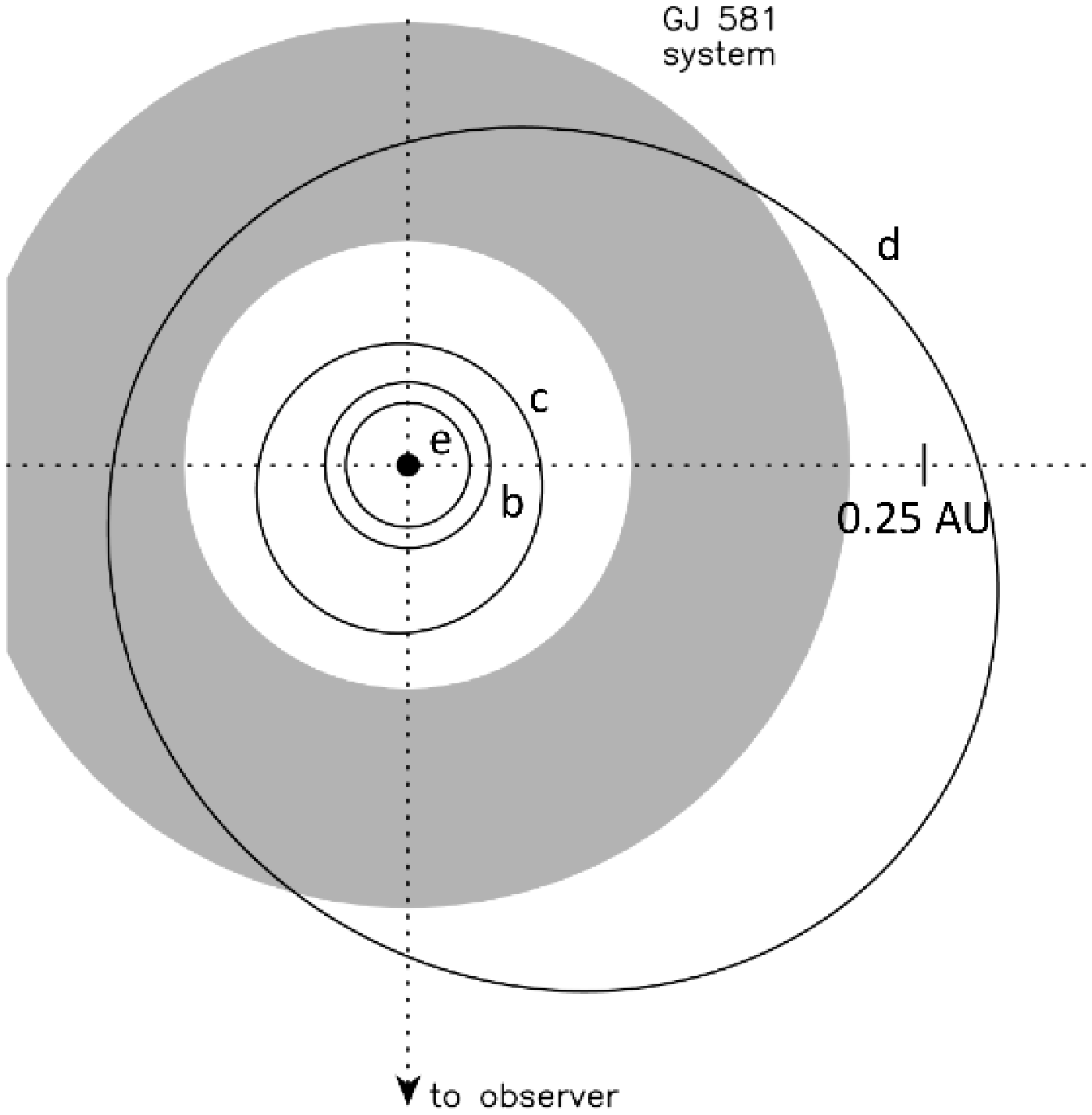}{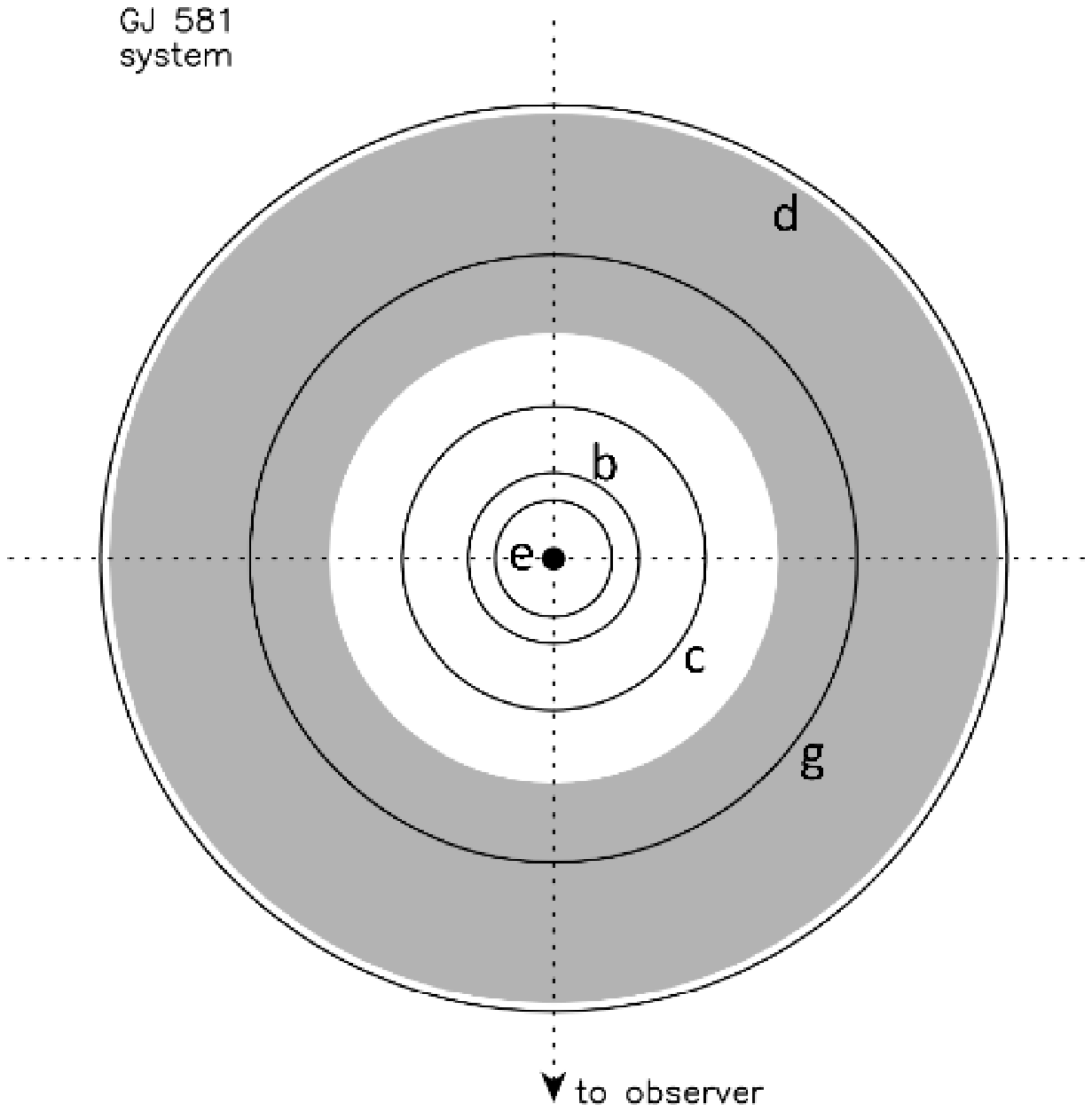}{fig_GJ581}{\
The HZ and planetary orbits of the GJ 581 system.
Left: Four planet scenario \citep{Mayor10}.
Planet d spends most of its 67-day orbit
in the HZ, thereby changing its equilibrium temperature by more than 80K.\
Right: Six planets scenario \citep{Vogt11}. Planet g's 36-day circular \
orbit is inside the HZ, while planet d's 67-day orbit puts it \
right at the HZ's outer edge. Any planet in the system's 
HZ with a moderately dense atmosphere could harbor liquid water}

\subsection{Are you Sirius? In and Out of the H$\alpha$ Absorption Line}

Limb darkening in absorption lines is expected to be less than it would be 
in the continuum because the line is formed all 
the way out to the stellar surface. More recent modeling shows that
limb darkening changes between the line wings and center also.
See Figure \ref{fig_Sirius}.

\articlefiguretwo{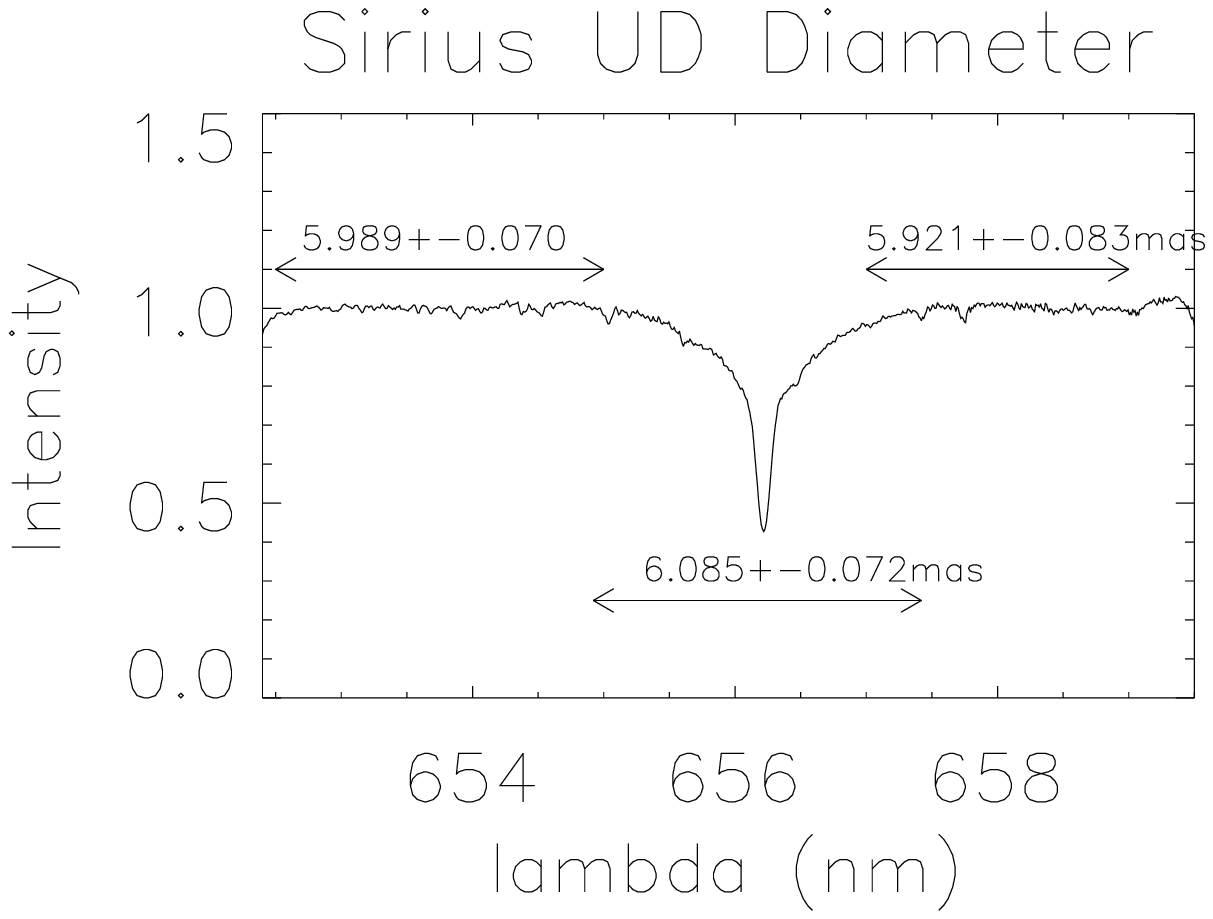}{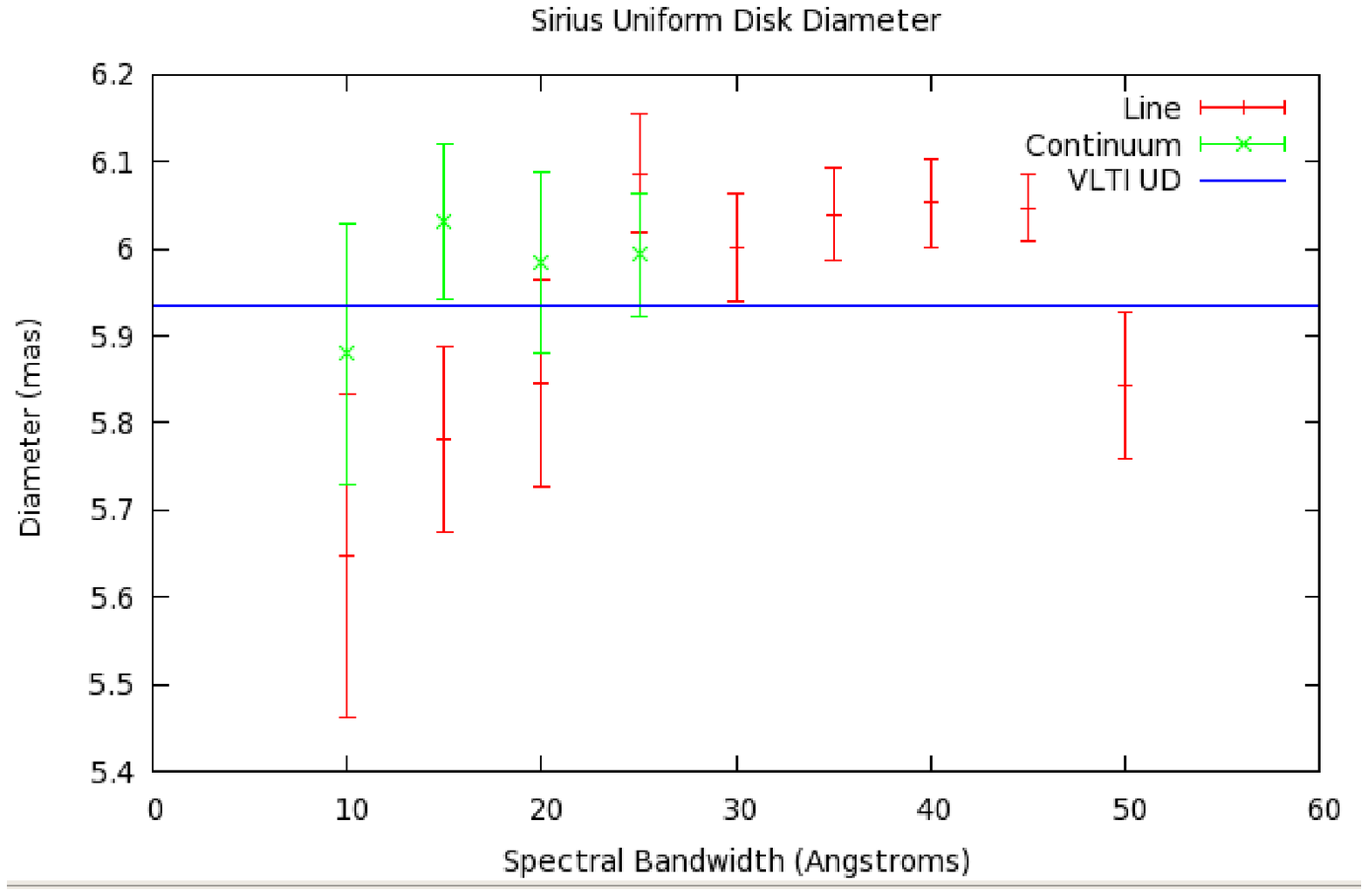}{fig_Sirius}{\
Left: Data taken by the VEGA beam combiner shows the uniform disk size 
inside the H$\alpha$ is indeed larger than on either side of the absorption 
line. Right: The same data analyzed with band-passes centered on the line 
and with a range of bandwidths. Here the uniform disk size seems to be smaller 
in the line center, while being larger once the line wings are included.}

\acknowledgements 
The CHARA Array is funded by the National Science
Foundation through NSF grant AST-0606958, by Georgia
State University through the College of Arts and Sciences, 
the W.M. Keck Foundation, The NASA Exoplanet Science Institute, and the
David and Lucile Packard Foundation.

\bibliography{ten_Brummelaar}

\end{document}